\documentclass[reprint,amsmath,amssymb,aps,prd]{revtex4-2}

\usepackage{graphicx}
\usepackage{dcolumn}
\usepackage{xcolor}
\usepackage{bm}
\usepackage{hyperref}

\newcommand{\drm}{{\rm d}}

\begin{document}

\title{Design and implementation of a seismic Newtonian-noise cancellation system for the Virgo gravitational-wave detector}

\author{Soumen Koley}
\email{soumen.koley@gssi.it}
\address{Gran Sasso Science Institute (GSSI), I-67100 L'Aquila, Italy}
\address{INFN, Laboratori Nazionali del Gran Sasso, I-67100 Assergi, Italy}
\author{Jan Harms}
\address{Gran Sasso Science Institute (GSSI), I-67100 L'Aquila, Italy}
\address{INFN, Laboratori Nazionali del Gran Sasso, I-67100 Assergi, Italy}
\author{Annalisa Allocca}
\address{Universit\`a di Napoli ``Federico II", I-80126 Napoli, Italy}
\address{INFN, Sezione di Napoli, I-80126 Napoli, Italy}
\author{Francesca Badaracco}
\address{INFN, Sezione di Genova, via Dodecaneso, I-16146 Genova, Italy}
\author{Alessandro Bertolini}
\address{Nikhef, 1098 XG Amsterdam, The Netherlands}
\author{Tomasz Bulik}
\address{Astronomical Observatory, University of Warsaw, Al. Ujazdowskie 4, 00-478 Warsaw, Poland}
\address{Nicolaus Copernicus Astronomical Center, Polish Academy of Sciences, ul. Bartycka 18, 00-716 Warsaw, Poland}
\author{Enrico Calloni}
\address{Universit\`a di Napoli ``Federico II", I-80126 Napoli, Italy}
\address{INFN, Sezione di Napoli, I-80126 Napoli, Italy}
\author{Marek Cieslar}
\address{Nicolaus Copernicus Astronomical Center, Polish Academy of Sciences, ul. Bartycka 18, 00-716 Warsaw, Poland}
\author{Rosario De Rosa}
\address{Universit\`a di Napoli ``Federico II", I-80126 Napoli, Italy}
\address{INFN, Sezione di Napoli, I-80126 Napoli, Italy}
\author{Luciano Errico}
\address{Universit\`a di Napoli ``Federico II", I-80126 Napoli, Italy}
\address{INFN, Sezione di Napoli, I-80126 Napoli, Italy}
\author{Marina Esposito}
\address{Universit\`a di Napoli ``Federico II", I-80126 Napoli, Italy}
\address{INFN, Sezione di Napoli, I-80126 Napoli, Italy}
\author{Irene Fiori}
\address{European Gravitational Observatory (EGO), I-56021 Cascina, Pisa, Italy}
\author{Stefan Hild}
\address{Maastricht University, 6200 MD Maastricht, The Netherlands}
\address{Nikhef, 1098 XG Amsterdam, The Netherlands}
\author{Bartosz Idzkowski}
\address{Astronomical Observatory, University of Warsaw, Al. Ujazdowskie 4, 00-478 Warsaw, Poland}
\author{Alain Masserot}
\address{Universit\'e Savoie Mont Blanc, CNRS, Laboratoire d’Annecy de Physique des Particules - IN2P3, F-74000 Annecy, France}
\author{Beno\^it Mours}
\address{Universit\'e de Strasbourg, CNRS, IPHC UMR 7178, F-67000 Strasbourg, France}
\author{Federico Paoletti}
\address{INFN, Sezione di Pisa, I-56127 Pisa, Italy}
\author{Andrea Paoli}
\address{European Gravitational Observatory (EGO), I-56021 Cascina, Pisa, Italy}
\author{Mateusz Pietrzak}
\address{Nicolaus Copernicus Astronomical Center, Polish Academy of Sciences, ul. Bartycka 18, 00-716 Warsaw, Poland}
\author{Luca Rei}
\address{INFN, Sezione di Genova, via
Dodecaneso, I-16146 Genova, Italy}
\author{Lo\"ic Rolland}
\address{Universit\'e Savoie Mont Blanc, CNRS, Laboratoire d’Annecy de Physique des Particules - IN2P3, F-74000 Annecy, France}
\author{Ayatri Singha}
\address{Maastricht University, 6200 MD Maastricht, The Netherlands}
\address{Nikhef, 1098 XG Amsterdam, The Netherlands}
\author{Mariusz Suchenek}
\address{Nicolaus Copernicus Astronomical Center, Polish Academy of Sciences, ul. Bartycka 18, 00-716 Warsaw, Poland}
\author{Maciej Suchinski}
\address{Astronomical Observatory, University of Warsaw, Al. Ujazdowskie 4, 00-478 Warsaw, Poland}
\author{Maria Concetta Tringali}
\address{European Gravitational Observatory (EGO), I-56021 Cascina, Pisa, Italy}
\author{Paolo Ruggi}
\address{European Gravitational Observatory (EGO), I-56021 Cascina, Pisa, Italy}
\date{\today}

\begin{abstract}
Terrestrial gravity perturbations caused by seismic fields produce the so-called Newtonian noise in gravitational-wave detectors, which is predicted to limit their sensitivity in the upcoming observing runs. In the past, this noise was seen as an infrastructural limitation, i.e., something that cannot be overcome without major investments to improve a detector's infrastructure. However, it is possible to have at least an indirect estimate of this noise by using the data from a large number of seismometers deployed around a detector's suspended test masses. The noise estimate can be subtracted from the gravitational-wave data; a process called Newtonian-noise cancellation (NNC). In this article, we present the design and implementation of the first NNC system at the Virgo detector as part of its AdV+ upgrade. It uses data from 110 vertical geophones deployed inside the Virgo buildings in optimized array configurations. We use a separate tiltmeter channel to test the pipeline in a proof-of-principle. The system has been running with good performance over months.
\end{abstract}

\maketitle

\section{Introduction}
The detection of gravitational waves (GWs) from a binary black hole coalescence in 2015 \citep{AbEA2016b} by the Advanced LIGO detectors \citep{LSC2015} marked the start of a new era in GW astrophysics. Ever since, the Advanced Virgo (AdV) and Advanced LIGO detectors \citep{AcEA2014,LSC2015} have detected more than a hundred GW signals \cite{AbEA2021a} spanning over three observing runs. Each of the observing runs were followed by a period of instrument upgrade and commissioning \citep{acernese2019advanced} aimed at improving the sensitivity and the duty-cycle of the detectors. The AdV detector achieved a binary neutron star range of about 60\,Mpc towards the end of the third observing run which lasted until March, 2020 \citep{acernese2022virgo}. Following this, a series of instrument upgrades were planned to achieve the Advanced Virgo Plus (AdV+) sensitivity \citep{flaminio2020status}. Design and implementation of an online Newtonian noise cancellation (NNC) system was one of the planned activities aimed at improving the low-frequency sensitivity of the detector during the first phase of AdV+ upgrades. From an astrophysical standpoint, improving the low-frequency sensitivity would increase the possibilities of detecting GW signals from stellar-mass black hole mergers and also increase the rate of detection of intermediate-mass binaries. Additionally, better constrained estimates of parameters like the chirp mass and effective spin of the binaries is also expected \citep{yu2018prospects}.
\par
Terrestrial gravity noise also known as Newtonian noise (NN) originates due to the gravitational coupling of ambient density fluctuations to the suspended test masses of the interferometer \citep{Sau1984}. These density fluctuations can be either atmospheric due to pressure and temperature fluctuations \citep{Cre2008,FiEA2018}, or of the subsurface due to the propagation of seismic waves \citep{BeEA1998,HuTh1998}. The former is referred to as atmospheric NN, while the latter is referred to as seismic NN. In this article, we address the cancellation strategies concerning seismic NN.
\par
\begin{figure}
\begin{center}
    \includegraphics[width=0.48\textwidth]{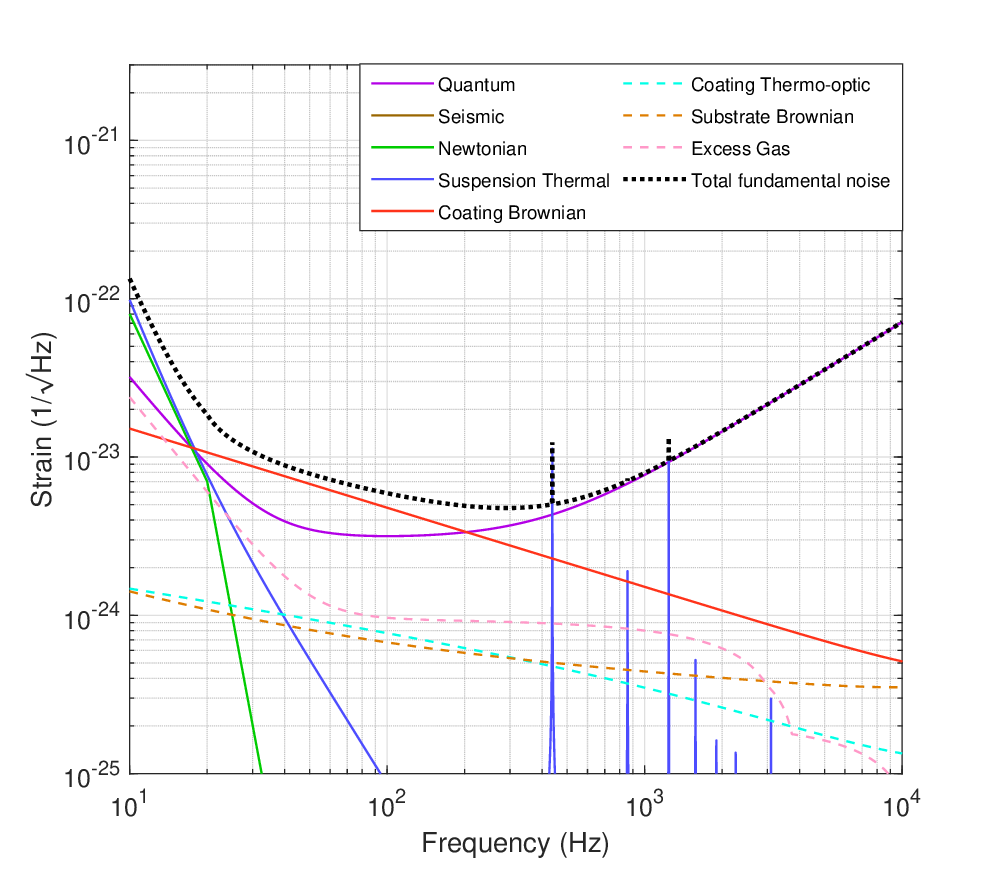}
    \caption{Contribution of several fundamental sources of noise to the AdV+ design sensitivity corresponding to a laser input power of 40\,W and 12\,dB of frequency-dependent squeezing. Newtonian noise is expected to be one of the major contributors to the low-frequency sensitivity.}
    \label{AdVPlusDesign}
\end{center}
\end{figure}

Newtonian noise is expected to be one of the major fundamental limits to the sensitivity of the AdV+ detector in the frequency band 10--20\,Hz. Figure \ref{AdVPlusDesign} shows the contribution of the several fundamental sources of noise to the AdV+ design sensitivity \citep{flaminio2020status}. The contribution of NN to the low-frequency sensitivity of the detector can be estimated by either using analytical models \citep{Har2019} or by finite element simulations of the seismic wavefield in the vicinity of the test-mass \citep{SHH2020,SiEA2021,baderThesis}. Both approaches require surface-seismic array studies aimed at deciphering the dominant wave type at the site (surface or body waves) and also quantifying the contribution of each of the wave-types from the different anthropogenic sources of noise. Prior to the design of the NNC for AdV+, several surface-seismic array studies have been conducted inside each of the end buildings \citep{TrEA2019, SiEA2021} and outside the interferometer arms \citep{koley2017s}.
\par
Based on the understanding of the propagation characteristics of the seismic waves near the test-masses of the interferometer, it is possible to design an optimal surface-array of seismometers for NNC. A first such NNC system was proposed in \cite{Cel2000} which makes use of the correlation between the ground motion measured by seismometers near the test-masses and the main interferometer signal. The underlying principle for NNC systems makes use of the linear relation between the measured ground motion and the expected Newtonian noise in order to design a Wiener filter corresponding to each of the seismometers \cite{CoEA2018a,HaEA2020}. Application of such a subtraction scheme was fully simulated in time domain \citep{DHA2012}. In cases when the NN originates due to a pure Rayleigh wavefield, studies by \citep{HaVe2016} have shown that NNC by even one tiltmeter would achieve NN residuals that would be limited only by the tiltmeter self-noise. The study also shows that for a more pessimistic scenario, when the seismic noise is a mixture of body and surface waves, a modest cancellation by about a factor two would be possible. However, before a noise-cancellation scheme can be implemented and tested, the positions of seismometers near the test-masses need to be determined for optimal NNC. Determination of the optimal locations of the seismometers for NNC is an optimization problem that minimizes a residual, which can be estimated by making use of the cross-correlations between seismometers and that between the seismometers and the expected Newtonian noise. For AdV+, based on prior estimates of correlations between seismometer channels, a Particle Swarm optimizer was used to determine the optimal geometry of the NNC arrays corresponding to each of the end buildings \citep{BaEA2020}.

In this paper, we present the results of the cancellation of the tilt signal measured at the North End Building (NEB) of the AdV+ detector \citep{allocca2021picoradiant} by using the seismic noise data measured by the NEB NNC array. In section \ref{sec:site}, we present the seismic wavefield characteristics at the NEB and prove that it is dominated by Rayleigh waves, a case in which the tilt signal can be used as a proxy for the expected NN. In Section \ref{sec:NNChar}, we present the NN estimates for the AdV+ detector based on array studies and finite element simulations. In Section \ref{sec:SysDesign} we present the optimization results that helped in designing the surface array of seismometers for the NNC system. In Section \ref{sec:wiener} we derive the expressions corresponding to the time-domain implementation of the Wiener filter for a Multiple-Input-Single-Output (MISO) system and detail the several signal processing steps implemented in the NNC pipeline. The noise-cancellation performance of the system when using the tilt signal as the target is presented in Section \ref{sec:principle}. Finally, we present the conclusions of our work in Section \ref{sec:conclusion}.

\section{Site characteristics}
\label{sec:site}
\begin{figure*}
\begin{center}
    \includegraphics[trim={0cm 0.65cm 0cm 0cm},clip,width=\textwidth]{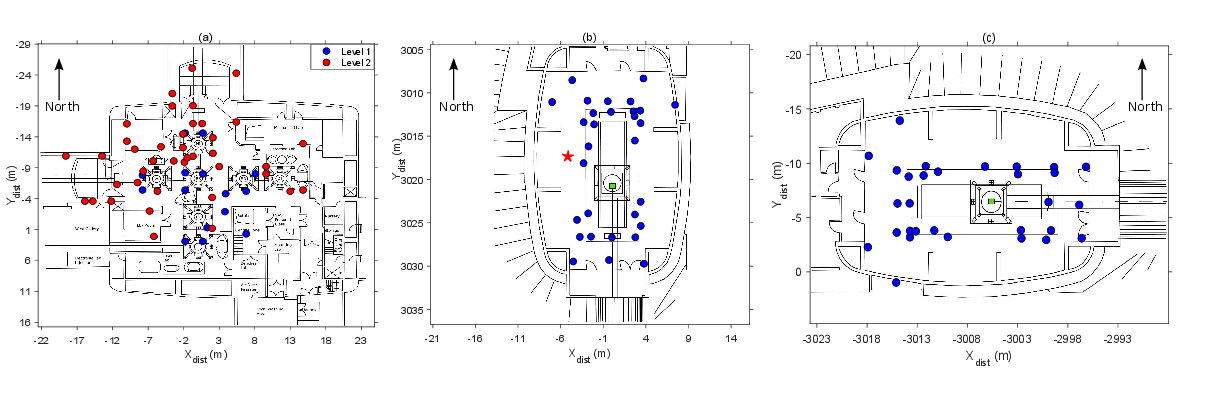}
    \caption{(a) The blue and red solid circles show the positions of the geophones at level 1 and 2 of the CEB, respectively. (b) The blue solid circles show the locations of the geophones and the red star shows the location of the tiltmeter in the NEB. (c) The blue solid circles show the locations of the geophones in the WEB. Note that the origin of the coordinate system corresponds to the location of the beamsplitter at the CEB, and `north' corresponds to the direction of the north arm of the interferometer which is oriented $20\circ$ clockwise with respect to the geographic north.}
    \label{NNCArray}
\end{center}
\end{figure*}

The AdV+ NNC array comprises a total of 110 seismic sensors, with 55 sensors deployed at the Central Building (CEB), and 30 sensors each at the NEB and the West End Building (WEB). These sensors were deployed in 2020 in their optimal positions (see Section \ref{sec:SysDesign}) with some refinements of the CEB array a year later. Each sensor is equipped with a vertical geophone with a resonance frequency of 4.5 Hz and a data acquisition system. Sensors creating an array are connected to an SPU (signal processing unit), providing sensor communication, time synchronization, and power. The sensor data acquisition system samples the seismic signal from the geophone at 500 samples/s and sends data after time synchronization to the storage server through the SPU. SPU provides sensors with a modified ethernet standard, which uses a communication speed of 100 Mb/s and only two pairs of ethernet cables. Another pair is used to provide time synchronization pulses generated every 1\,s. The last pair provides the power supply. These installations follow an initial site-characterization phase with deployments of temporary arrays in 2018 as reported in \cite{TrEA2019}. Figures \ref{NNCArray} (a), (b), and (c) show the positions of the sensors at the CEB, NEB, and WEB, respectively. In this Section, we present the amplitude and the phase characteristics of the seismic noise corresponding to the layout after the optimization studies were done.

The amplitude characteristics of the seismic noise data presented here were acquired between April 01, 2023 and May 07, 2023 at a sampling frequency of 500 Hz. Seismic data corresponding to each of the geophones are first divided into 1200 s long segments, and the instrument response is deconvolved, which applies a correction to the amplitude and the phase of the seismic data and converts it from voltage to ground velocity. The (single-sided) power spectral densities (PSDs) are then computed as
\begin{equation}
S_m(f) = 2|X_m(f)|^2/T,
\end{equation}
where $X_m(f)$ is the Fourier transform of the deconvolved seismic data at frequency $f$ of geophone $m$, and $T$ is the segment duration. The discrete Fourier transform is calculated using a Hamming spectral window \citep{oppenheim1999discrete}. The estimated spectral densities for every 1200\,s segment are then used to generate histograms with a bin size of 0.5\,dB ($\mathrm{1\,dB=20{log}_{10}(m/s/\sqrt{Hz})}$). Next, the $\mathrm{10^{th}}$, $\mathrm{50^{th}}$, and $\mathrm{90^{th}}$ percentile PSDs are extracted from the histogram. The process is then repeated for all the geophones in the NNC array.

\begin{figure*}
\begin{center}
    \includegraphics[trim={0cm 0.0cm 0cm 0cm},clip,width=\textwidth]{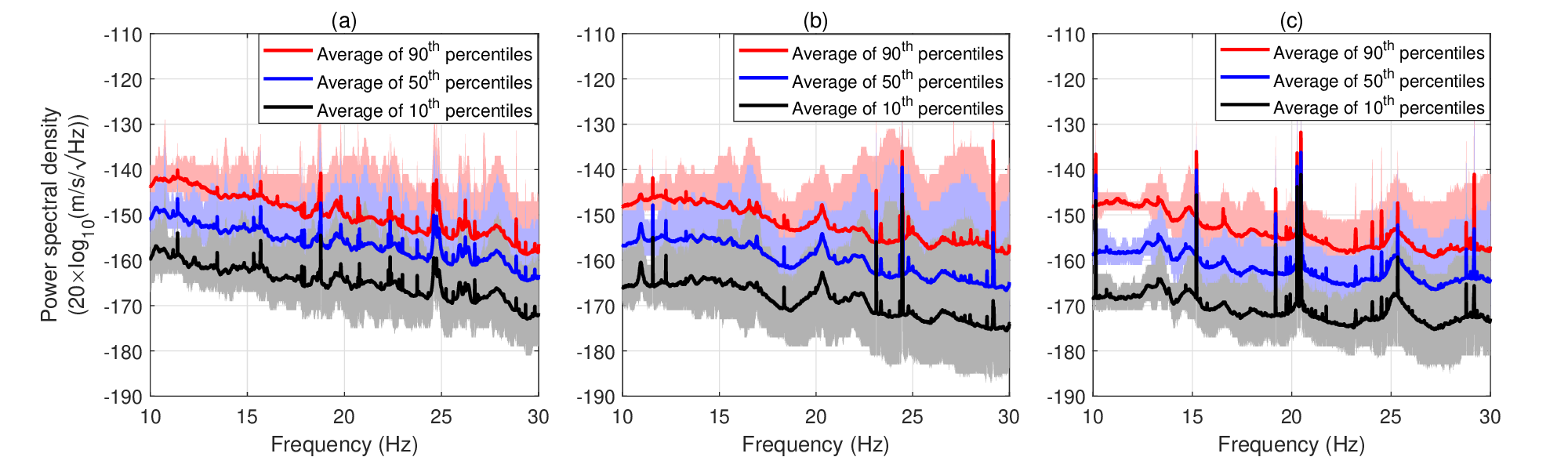}
    \caption{(a) The black, blue, and red shaded regions show the maximum and minimum of the $\mathrm{10^{th}}$, $\mathrm{50^{th}}$, and the $\mathrm{90^{th}}$ percentiles of the PSDs for the CEB geophones. The solid black, blue, and red curves show the average of the $\mathrm{10^{th}}$, $\mathrm{50^{th}}$, and $\mathrm{90^{th}}$ percentile PSDs. (b) and (c) are same as (a), but corresponds to the WEB and NEB, respectively.}
    \label{PSDAll}
\end{center}
\end{figure*}

The black, blue, and red shaded regions in Figure \ref{PSDAll}(a) show the maximum and minimum of the $\mathrm{10^{th}}$, $\mathrm{50^{th}}$, and the $\mathrm{90^{th}}$ percentiles of the PSDs for the CEB geophones. The solid black, blue, and red curves show the average of the $\mathrm{10^{th}}$, $\mathrm{50^{th}}$, and $\mathrm{90^{th}}$ percentile PSDs. Figures \ref{PSDAll}(b) and (c) show the same corresponding to the WEB and NEB, respectively. Seismic noise at each of the buildings vary between $3\times10^{-14}$ -- $10^{-18} \, \mathrm{m^2/s^2/Hz}$. A spatial variability of about 20\,dB is observed for frequencies between 10 -- 15\,Hz and it increases to about 40\,dB for frequencies above 20\,Hz. Besides broadband noise, several peaks are observed in the PSDs. These (nearly) monochromatic peaks are observed at the rotation frequency (or their harmonics) of the fans, motors, and pumps that constitute the heating, ventilation, and air conditioning system (HVAC) \citep{TringaliHVAC}. The HVAC system is necessary for operating the experimental equipments and the clean rooms near the test-masses.

The spatial variation of PSDs inside the end buildings can be used to point to sources of noise. However, it gives very limited information on the propagation characteristics of the noise. For that we use the phase difference between the noise measured at different geophones. Here we study the phase characteristics of the noise measured at the NEB, since we want to establish that the dominant wave type propagating at the NEB is the Rayleigh wave.

\begin{figure}
\begin{center}
    \includegraphics[width=0.45\textwidth]{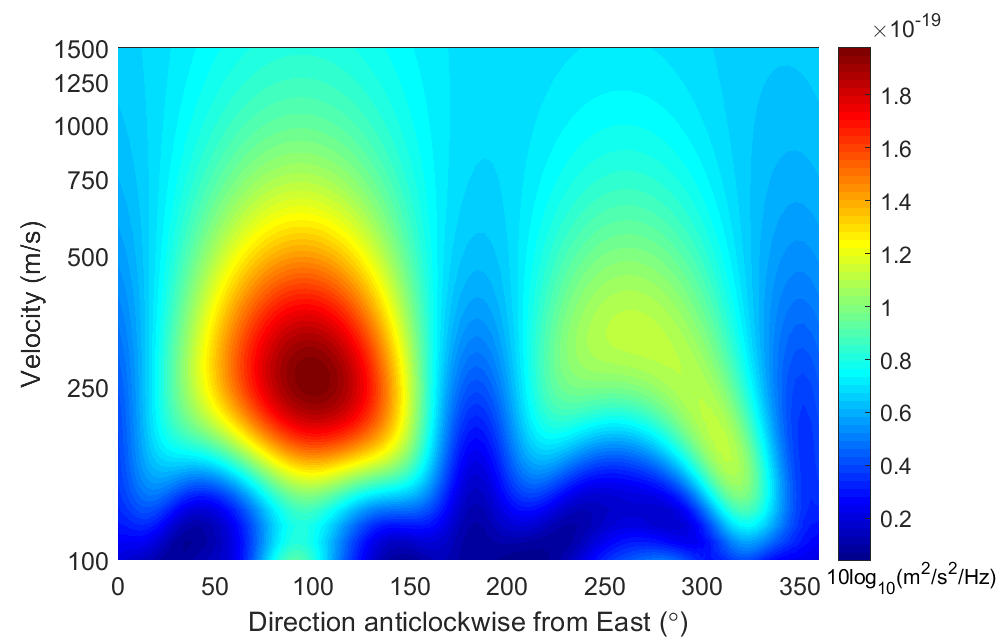}
    \caption{Frequency-domain beampower estimated using the NNC NEB array seismic noise data in the band 10--10.2\,Hz showing seismic noise propagating at speeds of about 250\,m/s and propagating from the north.}
    \label{NEBBP}
\end{center}
\end{figure}

The first metric we present is the propagation velocities of the seismic waves for different frequency bands. We use a plane wave beamformer \citep{lacoss1969estimation}, and estimate the frequency-domain beampower (FDB) $\mathrm{BP}$ as a function of slowness $p$ (inverse of speed) and direction of propagation $\phi$, which is measured anticlockwise from an eastward direction. The FDB for $N$ concurrent segments of seismic data measured at $M$ geophones is estimated as
\begin{equation}
    \mathrm{BP}(p,\phi) = \mathbf{w}(p,\phi,f)\mathbf{S}(f)\mathbf{w}^\dagger(p,\phi,f),
    \label{eq:beampower}
\end{equation}
where $\mathbf{S}(f)$ is the matrix of cross-spectral densities with $M\times M$ components
\begin{equation}
S_{m_1m_2}(f)=\frac{1}{N}\sum\limits_{n=1}^N2X^n_{m_1}(f)X_{m_2}^{n,\,*}(f)/T,
\label{eq:CrossCov}
\end{equation}
where `*' denotes the complex conjugate operator. The vector $\mathbf{w}(p,\phi,f)$ has $m=1,\ldots,M$ components $\exp(-2j\pi f\tau_m(p,\phi,f))$ representing the phase delays for a plane wave to reach geophone $m$, and $j=\sqrt{-1}$. The time delay for a plane wave can be expressed as $\tau_m(p,\phi,f) = x_mp \cos(\phi) + y_mp\sin(\phi)$, where $(x_m,y_m)$ are the coordinates of the $m^{\rm th}$ geophone. Following equation (\ref{eq:beampower}), we estimate the FDB for every 1200\,s stretch of data during the period April 01 -- May 07, 2023 corresponding to $v\in[100 ,1500]$\,m/s at an interval of 10\,m/s and $\phi \in [0^\circ,356^\circ]$ (measured anticlockwise from an eastward direction) at an interval of 4$^\circ$. We divide the 1200\,s of data into $N=12$ segments of length 100\,s, which means that the FDB is estimated with frequency bins of width 10\,mHz. An average over 20 adjacent bins is subsequently calculated to reduce the data volume. Figure \ref{NEBBP} shows the FDB averaged in the frequency band 10 -- 10.2\,Hz over all such 1200\,s windows during the entire measurement period. This band is characterized by a noise peak at 10.1\,Hz, which coincides with the rotation frequency of a fan, which is a part of the air handling unit (AHU) located in the technical room north of the NEB. The direction estimate from the FDB (Figure \ref{NEBBP}) matches with the location of the AHU, and the propagation speed of about 250\,m/s is evidence for a surface wave.

\begin{figure}
\begin{center}
    \includegraphics[width=0.45\textwidth]{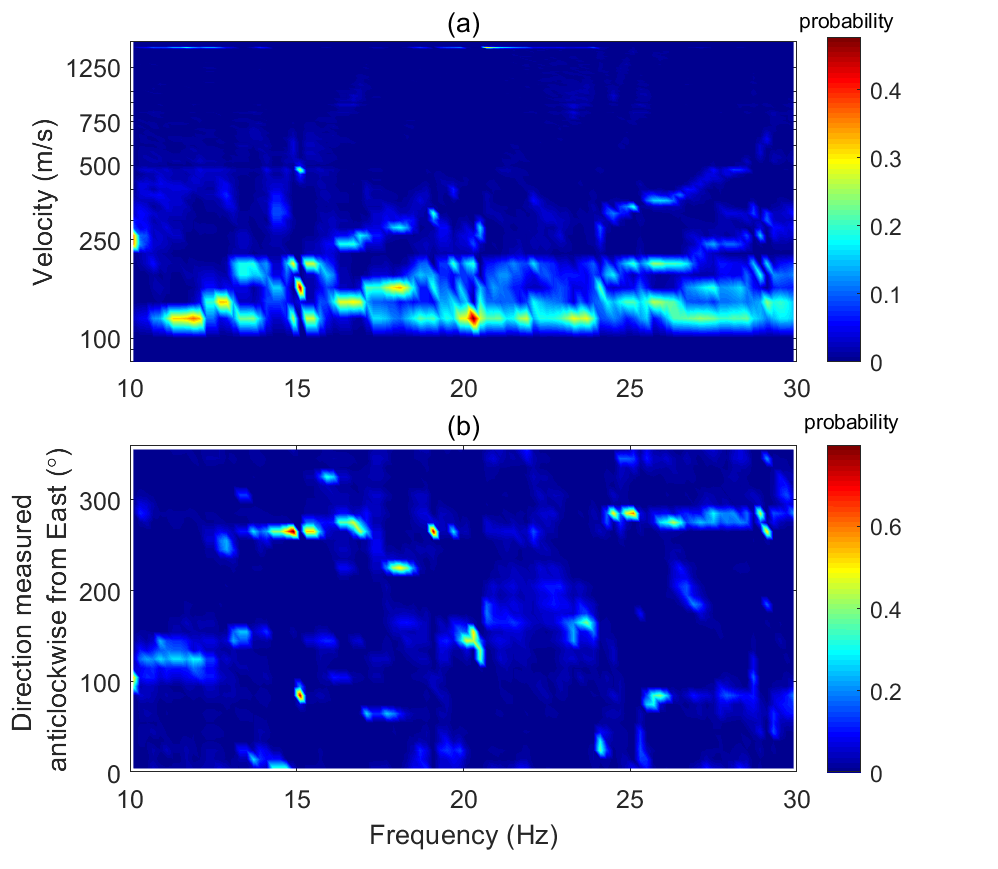}
    \caption{(a) Probability density functions of the speed (a) and direction (b) of propagation of seismic noise at the NEB corresponding to the frequency band 10 -- 30\,Hz at an interval of 0.2\,Hz.}
    \label{velPhiHist}
\end{center}
\end{figure}

In order to generate statistics about the wave-propagation attributes, the velocity and direction of propagation corresponding to the maximum FDB is stored for every 0.2\,Hz bin and histograms are generated using all 1200\,s windows. Later, we generate the probability density functions (PDFs) of the velocity and direction of propagation. The results are shown in Figures \ref{velPhiHist}(a) and (b). We observe that the seismic waves dominantly propagate with speeds between 100 -- 250\,m/s, which is characteristic of slowly propagating Rayleigh waves. The velocity PDFs for some frequency bands (Figure \ref{velPhiHist}(a)) show multiple peaks between 100 -- 250\,m/s and it corresponds to the different modes of Rayleigh wave propagation. Most of the noise originates either north or south of the array (Figure \ref{velPhiHist}(b)). For example, it is well known that the origin of the noise peaks at 10.1\,Hz, 15.2\,Hz, and 20.1\,Hz is the AHU located north of the NEB, and our analysis also points to the same direction. 

\par
\begin{figure}[ht!]
\begin{center}
    \includegraphics[width=0.45\textwidth]{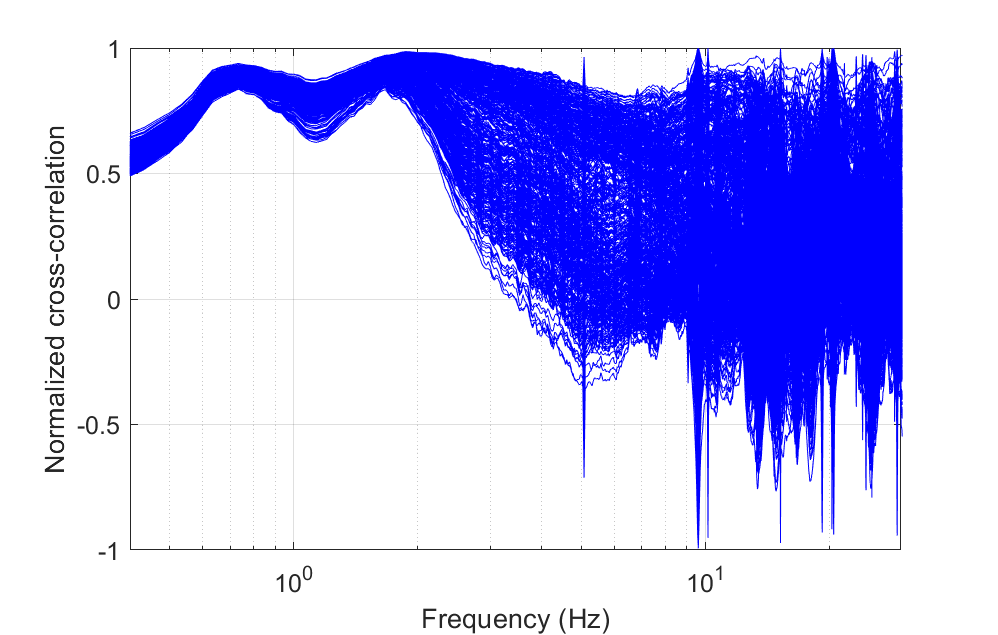}
    \caption{Normalized frequency domain cross-correlation corresponding to all the 435 station pairs at the NEB.}
    \label{NormCCNEB}
\end{center}
\end{figure}
\begin{figure}[ht!]
\begin{center}
    \includegraphics[width=0.45\textwidth]{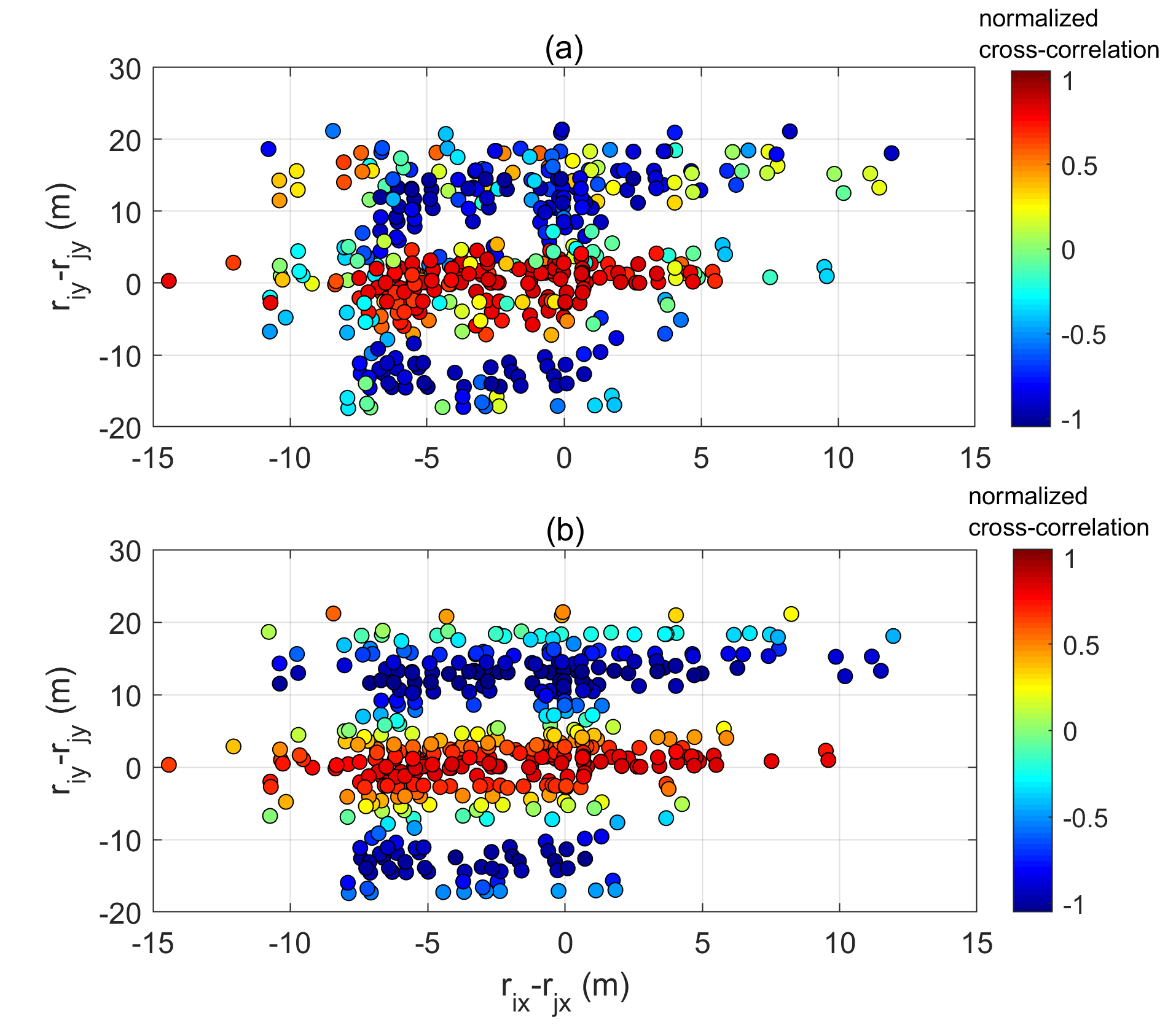}
    \caption{(a) Observed cross-correlations between geophone pairs at the NEB shown as a function of the relative position vector between them at a frequency of 10.1\,Hz. (b) Theoretical cross-correlations at a frequency of 10.1\,Hz computed using Eq. \ref{eqn4} corresponding to a velocity of 250\,m/s and direction of propagation between $80^\circ$ -- $100^\circ$.}
    \label{SpatialCC}
\end{center}
\end{figure}

The second metric we present are the normalized cross-correlations $C_{m_1m_2}(f)$ between geophones $m_1$ and $m_2$ which is computed as
\begin{equation}
C_{ij}(f)= \frac{\Re\left(S_{ij}(f)\right)}{\sqrt{S_{i}(f)S_{j}(f)}},
\label{eq:coherence}
\end{equation}
for each of the $N$ time segments, and $\Re$ represents the real part of a complex number. Figure \ref{NormCCNEB} shows the estimated cross-correlations between all 435 station pairs at the NEB. A strong positive correlation is observed for frequencies below 2\,Hz, since these frequencies are characterized by surface waves with large wavelengths \citep{koley2017s}. For the NNC frequency band between 10--30\,Hz, several peaks in the cross-correlation spectra are observed. These show both positive and negative cross-correlations and are characteristic of plane waves with wavelengths that are shorter than the array aperture ($\approx$ 25\,m for the NEB). These cross-correlations can be explained with an anisotropic plane wave (APW) model. The theoretical cross-correlation value $C^{\mathrm{APW}}_{i,j}(f)$ between geophones located at position vectors $\vec{r}_i$ and $\vec{r}_j$ corresponding to a propagation speed $v$ and angle of propagation between $\phi_1$ and $\phi_2$ is expressed as
\begin{multline}
C^{\mathrm{APW}}_{i,j}(f) = \frac{1}{A_P(f)}\int\limits_{\phi=\phi_1}^{\phi_2}\drm\phi\, A_{\mathrm{APW}}(\phi,f) \\
\cos\left(\frac{2\pi f}{v}(\vec{r}_i-\vec{r}_j)\cdot(\hat{x}\cos\phi + \hat{y}\sin\phi)\right),
\label{eqn4}
\end{multline}
where $A_{\mathrm{APW}}(\phi,f)$ is the amplitude as a function of source azimuth, and $\hat{x}$, $\hat{y}$ represent the unit vectors along the east and north directions, respectively. An amplitude normalization factor $A_P(f) = \int\limits_{\phi=\phi_1}^{\phi_2}\drm\phi\, A_{\mathrm{APW}}(\phi,f)$ is applied in order to set cross-correlation values in the range [-1,1]. Figure \ref{SpatialCC}(a) shows the observed cross-correlations at 10.1\,Hz as a function of the relative position vector between the geophones. An APW model that reproduces the observed cross-correlations using Eq. (\ref{eqn4}) with $v = 250$\,m/s and $\phi=[80^\circ,110^\circ]$ is shown in Figure \ref{SpatialCC}(b). 

Several frequency bands do not show the strong positive and negative cross-correlations, and can be explained as a mixture of an APW and a Gaussian correlation model. The Gaussian correlation model between stations with position vectors $\vec{r}_i$ and $\vec{r}_j$ can be expressed as
\begin{equation}
    C^{\mathrm{Gauss}}_{ij}(f) = A_{\mathrm{G}}(f)\exp\left(\frac{-\vert\vec{r}_i-\vec{r}_j\vert^2}{\sigma^2(v,f)}\right),
\end{equation}
where $\sigma(v,f)=\frac{v}{\pi f}$, $v$ is the speed of the propagating wave at frequency $f$, and $A_{\mathrm{G}}$ is the amplitude of the source. Gaussian correlation models are used to model the effect of sources of noise within the array and the effect of wave reflection and scattering, which suppress correlations over larger distances compared to APW models. An application of Gaussian correlation models to the Advanced LIGO seismic data has been shown in \citep{CoEA2016a}. 

\begin{figure}[ht!]
\begin{center}
    \includegraphics[width=0.45\textwidth]{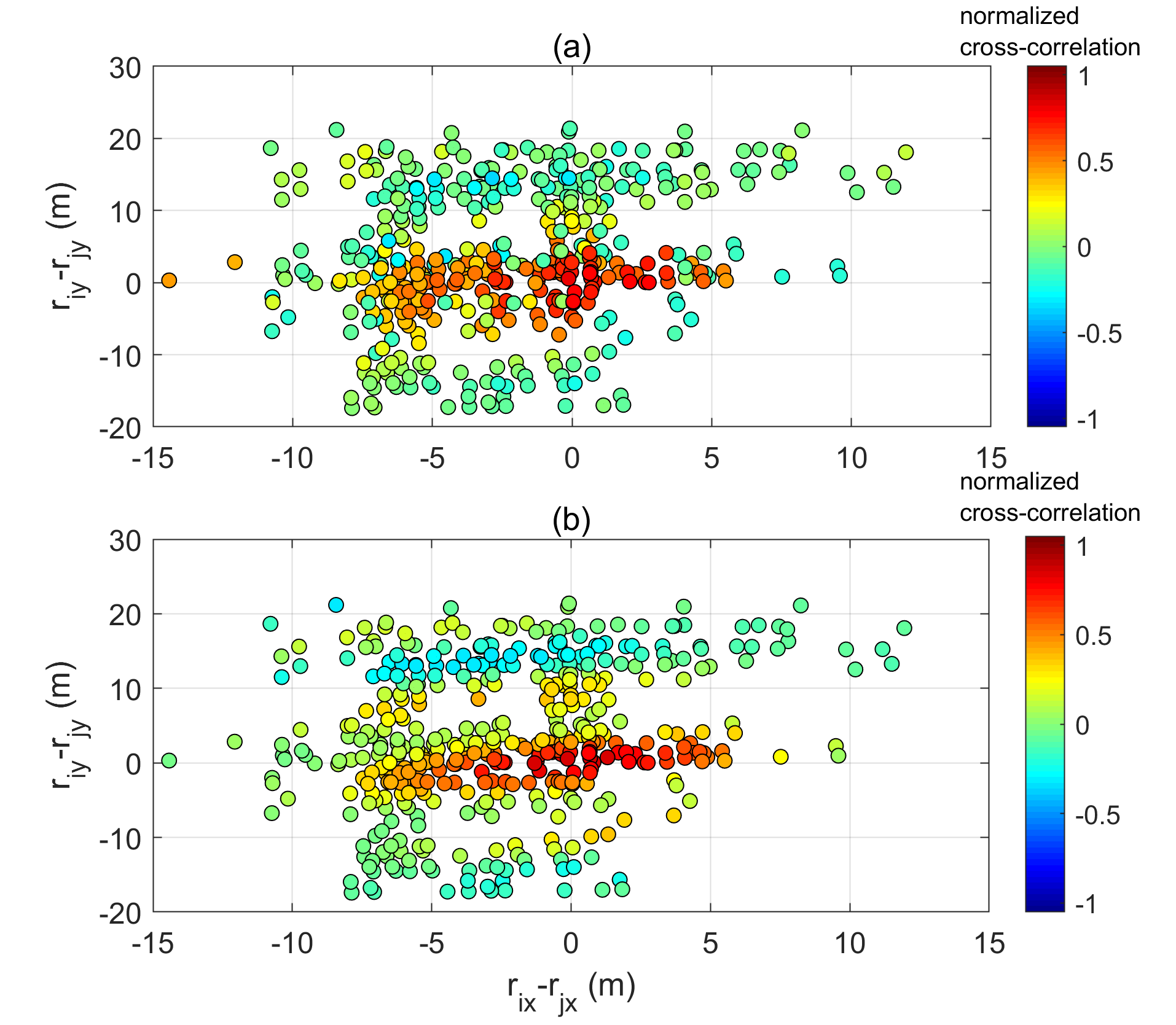}
    \caption{(a) Observed cross-correlations averaged in the frequency band 11--13\,Hz between geophone pairs at the NEB shown as a function of the position vector between them. (b) Theoretical cross-correlations computed using a mixture of APW and Gaussian correlation models corresponding to a speed of 110\,m/s and direction of propagation between $100^\circ$ - $130^\circ$.}
    \label{SpatialCCAPWGaussian}
\end{center}
\end{figure}

Figure \ref{SpatialCCAPWGaussian}(a) shows the spatial distribution of the observed cross-correlations averaged in the frequency band 11 -- 13\,Hz. We try to model the observed cross-correlations using a mixture of APW and Gaussian correlation models. Figure \ref{SpatialCCAPWGaussian}(b) shows the estimated cross-correlations corresponding to $v = 110$\,m/s and $\phi=[100^\circ,130^\circ]$. We observe that not all of the cross-correlations are reconstructed accurately using these models and is testament to the complexity of the seismic field inside the Virgo buildings.

In summary, the two metrics presented for interpreting the phase characteristics of the seismic noise at the NEB point to a dominant contribution from Rayleigh waves for the noise peaks. However, for the broadband noise, analytical models cannot reconstruct the observed correlations and only a part of the seismic field can be explained with a plane wave approach. Although we did not show any results from the CEB and WEB, seismic noise characteristics at these two buildings are similar to the NEB.

\section{Simulations of Virgo NN spectra}
\label{sec:NNChar}
The Virgo detector incorporates open spaces or recesses under the test masses as part of its clean room system. First calculations of Virgo's NN spectra relied on analytical equations that assume a flat surface, leaving room for improvement in accurately modelling the effects of recesses. The proper dimensions of the recesses under the input and end test mirrors in Virgo's central and end buildings were taken into account for NN estimation in \cite{SHH2020,SiEA2021}. Here we summarize the main results of these studies.

To assess the impact of the recesses, simulations were performed of an isotropic distribution of Rayleigh-wave propagation directions in the vicinity of the test-masses. The speed of Rayleigh waves is an important parameter, which was taken from an analysis similar to what is shown Figure \ref{velPhiHist}(a). The slower the waves (the shorter the wave length) the more effective the recess to reduce NN. Using a finite-element model, the resulting gravity perturbation caused by these Rayleigh waves was integrated. The mathematical formulation and parameters of the Rayleigh-wave field used in the simulation can be found in \cite{Har2019}. 

Simulations were done for frequencies between 5\,Hz and 25\,Hz. The integration over finite-element displacements, which gives rise to gravity perturbations $\delta a(\vec r_0, t)$ at the position $\vec r_0$ of the test-mass, can be expressed as follows: 
\begin{equation}
\begin{split}
     \delta a(\vec r_0, t) &= G\rho_0 \sum_{i}V_i \frac{1}{|\vec{r}_i - \vec{r}_0|^3}\\
     &\qquad\cdot\bigg(\vec{\xi}(\vec{r}_i, t) - 3(\vec{e}_i\cdot \vec{\xi}(\vec{r}_i, t))\cdot\vec{e}_i\bigg).
\end{split}
\label{eq:dipoleNN}
\end{equation}
In this equation, $\vec{r}_i$ represents the position of the $i^{\rm th}$ finite element of volume $V_i$, $\vec{\xi}(\vec{r}_i, t)$ is its seismic displacement, and $\vec{e}_i$ is the unit vector pointing along $\vec{r}_i-\vec{r}_0$. The finite-element model allows us to consider both the gravity perturbations resulting from vertical surface displacement and the compression or decompression of the underlying ground medium by summing over these effects. 

\begin{figure}[ht!]
    \centering
    \includegraphics[width=1.00\columnwidth]{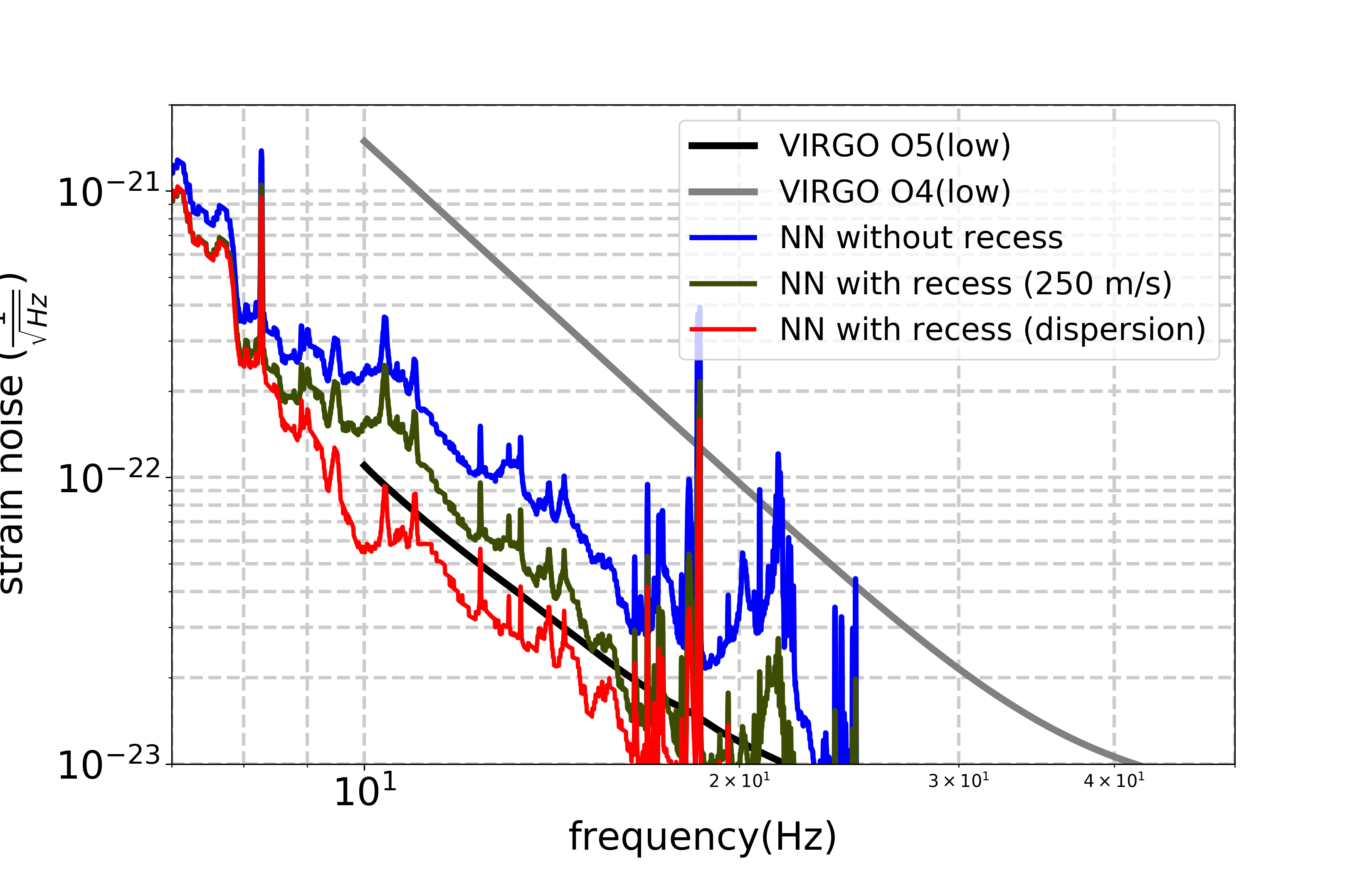}
  \caption{Comparison of Virgo NN estimates summing contributions from all four test masses as published in \cite{SiEA2021}. For reference, the low-noise models of the O4 and O5 observation runs are shown  \cite{Abbott2020x}. The blue curve represents the NN for a flat surface. The NN estimate with constant velocity of 250\,m/s is shown in green. The NN estimate considering the observed Rayleigh-wave dispersion is shown in red.}
  \label{fig:NN_overall}
\end{figure}

In the estimation of the total NN with contributions from four test masses, as presented in Figure \ref{fig:NN_overall}, the assumption was made that the NN in the 5--25\,Hz band is uncorrelated between the test masses. This assumption is certainly valid for NN correlations between the two test masses of an interferometer arm, but it might not be valid between the two 3\,km distant test masses inside the CEB. Compared to earlier analyses, it was found that the recess causes NN to be reduced by about a factor of 4 within the frequency range 10--20\,Hz when accounting for the observed Rayleigh-wave dispersion.

Accordingly, seismic NN is expected to largely fall below the sensitivity targets set for the upcoming observation runs, O4 and O5, with the exception of a few peaks. However, this analysis does not consider the contribution of NN transients associated with stronger transient waves of the seismic field. In fact, given the highly non-stationary character of the seismic field, one should expect frequent perturbation of Virgo data by NN transients in O5 \citep{AlEA2021}. 

\section{System design}
\label{sec:SysDesign}
The main goals of a NNC system design are to reduce NN in the GW detector data to an acceptable level and to do so reliably over months and years of operation. NNC systems will typically require a large number of sensors, which means that sensor failures and problems with data quality, e.g., introduced by electromagnetic disturbances or by the data-acquisition system, must be extremely rare. A single sensor whose data quality degrades for some reason while the NNC system is operational can spoil the NNC performance.

Concerning the NNC performance, as long as the goal is to reduce the NN spectral density (instead of subtracting NN transients, which is a different problem not specifically addressed with the current Virgo NNC design), it is determined by the correlations between all the seismometers and between seismometers and the Virgo GW channel, and by the signal-to-noise ratios of the seismic measurements. The correlations depend on what is actually measured, e.g., horizontal or vertical seismic displacement. If NN is not (yet) observed, the correlations between seismometers and GW detector data must be modeled. We describe below how these correlations are used to calculate optimal array configurations.

Finally, it must be decided what type of noise-cancellation filter is used. In section \ref{sec:principle}, we will present results for a time-invariant, finite-impulse response (FIR) Wiener filter, but adaptive Wiener filters, Kalman filters and other types of time-variant filters can be considered. In fact, we will argue in section \ref{sec:principle} and \ref{sec:discussion} that adaptive Wiener filters should not only be expected to achieve better average NNC performance, but also to solve practical issues.

\subsection{Instrumentation}
Assuming that seismic NN is dominated by contributions from Rayleigh waves, the natural choice for efficient NNC would be to monitor seismic displacements along the horizontal directions of the interferometer arms since this would lead to higher correlations between seismometers and seismic NN compared to vertical seismometers \cite{HaVe2016}. However, the seismic sensors deployed for the NNC system at Virgo are vertical geophones. The reason is that while our array measurements presented in section \ref{sec:site} provide strong evidence that Rayleigh waves make the dominant contribution to vertical surface displacement and seismic NN, horizontal surface displacement is expected to have significant contributions from Love waves. Love waves can only produce NN through inhomogeneous geology and non-planar surfaces, which means that in the presence of a dominant Rayleigh-wave field, the main effect of Love waves is to reduce correlations between horizontal seismometers and seismic NN. It is an immense benefit for NNC to deal with only one type of seismic wave \cite{BaHa2019}, and so the choice of vertical geophones measuring Rayleigh-wave displacement is justified. As a note, a tiltmeter was deployed at the NEB to investigate a potential utilization for NNC \cite{HaVe2016,allocca2021picoradiant}. It must be emphasized that the discussion so far is accurate only if the surface is flat, which is not the case at Virgo around its test masses \cite{SHH2020}. It has never been analyzed whether a combination of horizontal and vertical sensors or tiltmeters might lead to improved performance of the Virgo NNC system. 

Another technical design choice was to digitize the geophone data at the sensors, and to send the digitized data to a central data-acquisition unit. The rationale for this decision was that it avoids excess noise from ambient electromagnetic fluctuations coupling into cables and connectors transmitting the seismic data. This design comes with its own risks. For example, the timing and digitization of data at the sensors is a source of noise, and in fact, during the commissioning of the arrays, data-quality issues were noticed and eventually solved by modifying the sensor housing (increasing distance and adding EM shielding between geophone and digitizer). Also, receiving and packaging timed digitized data from many sensors is a complex operation, which can fail. Issues with this operation were identified in the early phase of the commissioning of the NNC system and had to be solved. The only remaining known data-quality issue is coming from loss of a few samples per day created by the digitizers of the individual sensors. However, these sample losses have a negligible impact on noise-cancellation performance.

\subsection{Optimizing the array configuration}
Virgo presents a complicated structure: the ground is not homogeneous and there is a basement under each test mass whose floor is 3.5\,m below the surface. At the end buildings, the walls of the basement are disconnected by a thin gap (5\,cm) from the main building floor, which reduces the transmission of external seismic disturbances \cite{TrEA2019}. The entire structure supported by the basement is called tower platform and it is anchored with 52\,m deep pillars to a more stable gravel layer beneath the clay (there are many gravel layers, which alternate with clay in the substrate of soil beneath Virgo \cite{DellaRocca1987studio}). These pillars are meant to prevent the sinking of the basement. This complex structure and the presence of local seismic sources entail a seismic field that is not describable with analytical models. Therefore, unlike LIGO, finding the optimal array to cancel NN in Virgo is not a trivial task and it is necessary to rely on measured seismic correlations \cite{CoEA2016a}. Correlation measurements can only be done between a finite set of points on the surface, and the full correlation function between any two points of the seismic field needs to be properly reconstructed. To search for the optimized array configuration two things are necessary: the reconstructed correlation function between seismometers and the correlation vector between seismometers and GW detector noise. The latter can be either modeled or measured.

In the following, we summarize the main results of \cite{BaEA2020}. Any optimization needs a cost function to be minimized. For NNC, a commonly used cost function is the spectral density of the residual noise $E(f)$ left after NN subtraction in the GW data. Expressed as a relative reduction of the NN spectral density $N(f)$, the cost function can be written as \cite{Cel2000}:
\begin{equation}\label{eq:res}
    \mathcal R(f) \equiv \frac{E(f)}{N(f)} = 1 - \frac{\mathbf{P}^{\dag}(f)\mathbf{S}^{-1}(f)\mathbf{P}(f)}{N(f)},
\end{equation}
where $\mathbf{S}(f)$ is the seismometer cross-power spectral density matrix of the seimometers and $\mathbf{P}(f)$ is the cross-power spectral density vector of seismometers and seismic NN.

The optimization can be performed by minimizing $- \mathbf{P}^{\dag}(\omega)\mathbf{S}^{-1}(\omega)\mathbf{P}(\omega)$, which is the term depending on the positions of the seismometers. The optimization for the NNC has to find the global minimum of $\mathcal R(f)$ with respect to the seismometer positions. This can be done with a stochastic optimization algorithm, such as Particle Swarm or a Genetic Algorithm.  At each step of the stochastic optimization, the value of the cost function (the residual) relative to randomly sampled array configuration is evaluated. The next configuration is then chosen following some criteria, which depend on the chosen algorithm \cite{Wales1997, Das2011, Bonyadi2017}. This means that the optimizer must be able to calculate the residual, and therefore $\mathbf{S}(f)$ and $\mathbf{P}(\omega)$, for any possible array configuration. 

Site-characterization measurements provide correlations only between a finite set of seismometer positions. Some form of interpolation of the correlation measurements is needed to carry out an array optimization. Standard interpolation techniques (linear, cubic, spline) are not accurate enough and in any case cannot be used to extrapolate to sensor positions outside the convex hull of the site-characterization array. More sophisticated Bayesian methods are computationally very expensive. A solution to this problem was presented in \cite{BaEA2020}. Instead of performing an interpolation of measured correlations $S(x_i,y_i,x_j,y_j, f)$ between sensors $i,\,j$, it is possible to interpolate the Fourier transform of the signals recorded by all seismic sensors, which only depends on two coordinates. It is then possible to evaluate $S(x_i,y_i,x_j,y_j, f)$ for any pair of sensor positions by exploiting the convolution theorem (see \cite{BaEA2020} for further details). To further accelerate the optimization process, one first evaluates $S(x_i,y_i,x_j,y_j, f)$ on a denser grid with the method discussed above, and then uses a standard interpolation technique for a rapid evaluation of $S(x_i,y_i,x_j,y_j, f)$ for arbitrary sensor locations. One thereby obtains a surrogate model of seismic correlations, which are also required for a model of $\mathbf{P}(\omega)$ \cite{Har2019}. This means that the full cost function $\mathcal R(f)$ is now given as a surrogate model and optimization can be performed. The results of such an optimization are shown in Figure \ref{NNCArray}. The optimization is performed at a specific frequency for an arbitrary (but fixed during optimization) number of sensors. It is also possible to perform an array optimization for broadband NNC. This can be done by building a cost function, for example, as a sum of residuals at different frequencies or by minimizing the maximum of residuals at different frequencies.

\section{Wiener filtering}
\label{sec:wiener}
The filtering of seismic data for NNC in the time domain is formulated as a MISO system where the multiple inputs comprise the data from the geophones (reference channels) and the target is the GW channel of the Virgo detector, or in our case, for test purposes, the tiltmeter signal measured at the NEB (see section \ref{sec:principle}). Given the linear relation between the measured seismic data and the expected NN and since the cost function is quadratic in the residuals, the objective is to estimate the optimal linear filter, i.e., Wiener filter, mapping samples of the geophones to a combined NN estimate. Following the Wiener theory \citep{levinson1946wiener}, the $k^{\rm th}$ sample of the error signal in the time domain is expressed as,
\begin{equation}\label{eq:error_funct}
    e(k) = y(k) - \mathbf{h}(k)\mathbf{x}(k),
\end{equation}
where $y(k)$ is the target signal, $\mathbf{h}(k) = [\mathbf{h_1}(k),\mathbf{h_2}(k),\cdots,\mathbf{h_M}(k)]_{(1\times ML)}$ is a row vector of the $M$ impulse responses each of length $L$, and $\mathbf{x}(k) = [\mathbf{x_1}(k),\mathbf{x_2}(k),\cdots,\mathbf{x_M}(k)]^\top_{(ML\times 1)}$ is a column vector of the past $L$ samples of the data measured at each of the $M$ reference channels, and $\top$ is the transpose sign. Hence, every element of the vector $\mathbf{x}(k)$ is a column vector of the form $\mathbf{x_m}(k) = [x_m(k),x_m(k-1),...,x_m(k-L+1)]^\top_{(L\times1)}$. Similarly, every element of $\mathbf{h}(k)$ can be expanded as $\mathbf{h_m}(k) = [h_m(0),h_m(1),\cdots,h_m(L-1)]_{(1\times L)}$. Thus, given the past $L$ samples of the reference data, the optimal impulse response per reference channel can be used to estimate the present sample of the target signal. The optimal set of impulse responses is obtained by solving $\partial E\{e^2(k)\}/\partial \mathbf{h}(k) = 0_{(ML\times 1)}$, and it yields
\begin{equation}
    \mathbf{h}(k) = \mathbf{P}\mathbf{S}^{-1}.
    \label{eqn9}
\end{equation}
The matrix $\mathbf{S}$ and row vector $\mathbf{P}$ can be expressed as
\begin{equation}
\mathbf{S} = 
\begin{pmatrix}
\mathbf{\Phi}_{11} & \mathbf{\Phi}_{12} & \cdots & \mathbf{\Phi}_{1M} \\
\mathbf{\Phi}_{21} & \mathbf{\Phi}_{22} & \cdots & \mathbf{\Phi}_{2M} \\
\vdots  & \vdots  & \ddots & \vdots  \\
\mathbf{\Phi}_{M1} & \mathbf{\Phi}_{M2} & \cdots & \mathbf{\Phi}_{MM}
\end{pmatrix},\;
\mathrm{and}\;\mathbf{P} = 
\begin{pmatrix}
    \mathbf{\Psi}_{y1}\\
    \mathbf{\Psi}_{y2}\\
    \vdots\\
    \mathbf{\Psi}_{yM}\\
\end{pmatrix}^\top.
\end{equation}
Each submatrix $\mathbf{\Phi}_{ij}$ of the block matrix $\mathbf{S}$ can be further written as,
\begin{equation}
\mathbf{\Phi}_{ij} = \begin{pmatrix}
  c_{ij}(0) & c_{ij}(1) & \cdots & c_{ij}(L-1)\\ 
  c_{ij}(-1) & c_{ij}(0) & \cdots & c_{ij}(L-2) \\
  \vdots & \vdots & \ddots & \vdots \\
  c_{ij}(1-L) & c_{ij}(2-L) & \cdots & c_{ij}(0)
\end{pmatrix},    
\end{equation}
where $c_{ij}(\tau)$ is the cross-correlation between the reference data measured at the $i^{\rm th}$ and $j^{\rm th}$ channels corresponding to the time lag $\tau$. Each element $\mathbf{\Psi}_{ym}$ of the row vector $\mathbf{P}$ can be expanded as, $\mathbf{\Psi}_{ym} = [c_{ym}(0),c_{ym}(1), \cdots, c_{ym}(L-1)]$. It should be noted that the cross-correlations $c_{ij}$ and $c_{ym}$ used in the Wiener filter calculation are typically averaged over a day of data. This makes the cross-correlations less sensitive to the temporal variability of the seismic data, and the performance of the Wiener filter becomes more robust.

The NNC system aims at removing the contribution of seismic NN in the frequency band 10--30\,Hz. Hence, for the real-time implementation, several stages of signal preconditioning are implemented to each of the reference channels before the Wiener output $\hat{y}(k) = \mathbf{h}(k)\mathbf{x}(k)$ can be subtracted from the target channel. At the first stage, the reference data acquired at a sampling frequency of $f_{\rm R}=500$\,Hz are decimated to $f_{\rm D}=100$\,Hz by using a Hamming window FIR (Finite Impulse Response) low-pass filter of order $M_{\rm A}=100$ and stopband frequency $f_{\rm S}\geq 35$\,Hz. It is important to note that a $M^{\rm th}$ order FIR filter has $(M+1)$ coefficients. At the second stage, the 100\,Hz reference data are high-pass filtered using a Hamming window FIR filter of order $M_{\rm B}=50$ and passband frequency $f_{\rm p}\geq 10$\,Hz. At the third stage, the Wiener filter of order $L = 100$ is applied to the low-pass and high-pass filtered data. However, before the Wiener output can be subtracted from the target data, it needs to be upsampled to the sampling frequency of the target data $f_{\rm T} = 10$\,kHz (for the tiltmeter signal) or 20\,kHz (for the Virgo strain signal). At the final stage, in order to remove the aliasing effect in the upsampled data, the data is low-pass filtered with a Hamming window FIR filter of order $M_{\rm C}=5000$ and $f_{\rm S} \geq 35$\,Hz. The upsampled Wiener output is finally subtracted from the target data and we produce the NN cancelled target data.
\begin{figure}
\centering
    \includegraphics[trim={0.7cm 0.0cm 1.4cm 0cm},clip,width=\columnwidth]{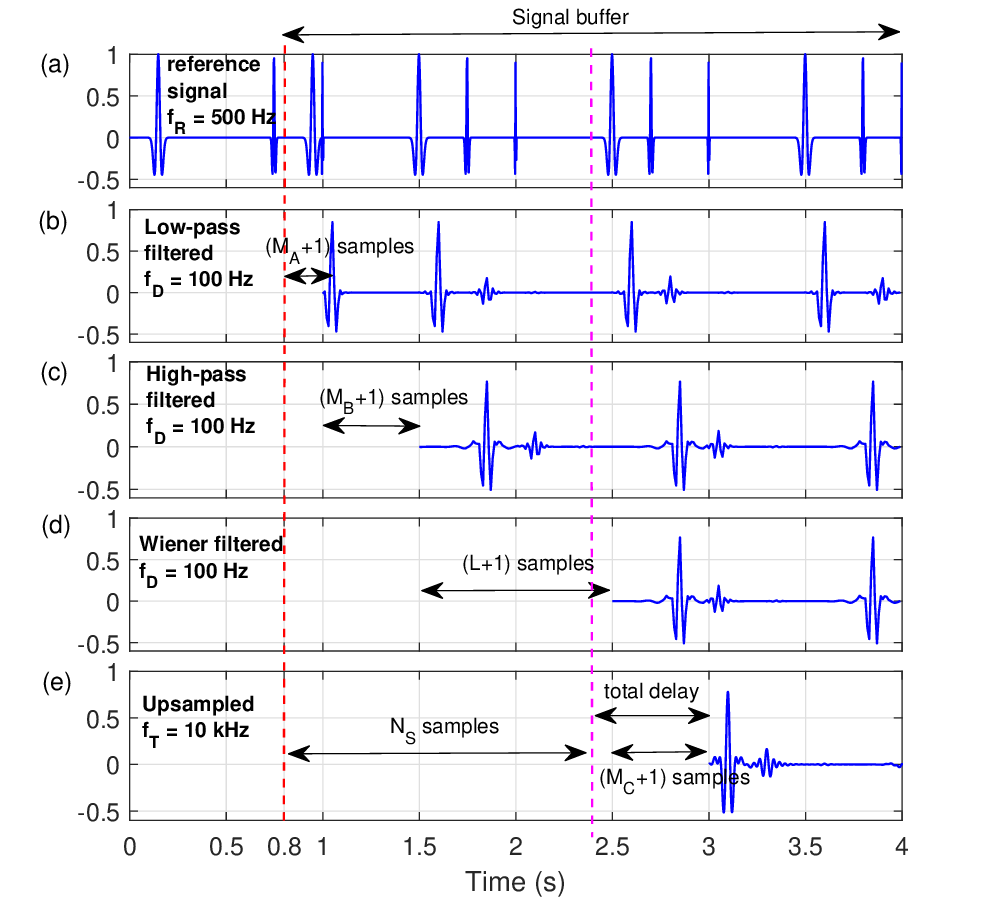}
    \caption{(a) A test signal sampled at 500\,Hz composed of Ricker wavelets with peak frequencies 20\,Hz, 40\,Hz, and 150\,Hz. The double arrow shows the length of the buffer. (b) Low-pass filtered signal with a Hamming window FIR filter of order $M_{\rm A}$ = 100, $f_{\rm S} \geq$ 35\,Hz, and decimated to 100\,Hz. (c) High-pass filtered signal with a Hamming window FIR filter of order $M_{\rm B}$ = 50, and $f_{\rm p} \geq$ 10\,Hz. (d) Wiener filtered signal with $L$ = 100. (e) Wiener output upsampled to 10\,kHz and low-pass filtered using a a Hamming window filter of order $M_{\rm C}$ = 5000, and $f_{\rm S} \geq$ 35\,Hz. At each stage of processing, the signals are delayed and reduced in length which equals the filter order.}
    \label{ToyFIR}
\end{figure}
\par
For the real-time implementation of the above-mentioned steps, two things must be addressed. Firstly, a circular buffer of the reference data needs to be maintained. This is due to the fact that the application of a FIR filter of order $M$ to a time series data produces the filtered output starting at the $(M+1)^{\rm th}$ sample of the data. Secondly, an FIR filter of order $M$ introduces a time delay of ($M/2$) samples ($M \in$ even positive integer), hence the Wiener output needs to be aligned in time with the target data before subtraction. The NNC application acquires data from the Virgo data stream every second and creates a buffer of $N_{\rm B}$ samples per channel. The length of this buffer can be expressed as
\begin{equation}
    N_{\rm B} = \left(\frac{M_{\rm A}}{f_{\rm R}} + \frac{M_{\rm B}}{f_{\rm D}} + \frac{L}{f_{\rm D}} + \frac{M_{\rm C}}{f_{\rm T}} + 1\right)f_{\rm R}.
    \label{eqn11}
\end{equation}
Corresponding to $f_{\rm R}$ = 500\,Hz, $f_{\rm T}$ = 10\,kHz, and the filter orders mentioned previously, a buffer of 1600 samples or 3.2\,s (corresponding to $f_{\rm R} = 500\,$Hz) is necessary. Hence, the first second of output is only produced at the end of the first 4\,s. The process then repeats and the buffer is replenished with new data every second. Next, we align the starting point of the one-second long Wiener output which was obtained by processing the 3.2\,s of the reference data. The starting sample $N_{\rm S}$ of the 3.2\,s long target data that aligns with the first sample of the upsampled Wiener output is expressed as
\begin{equation}
    N_{\rm S} = \left(\frac{M_{\rm A}f_{\rm D}}{2f_{\rm R}} + \frac{M_{\rm B}}{2} + L\right)\frac{f_{\rm T}}{f_{\rm D}} + \frac{M_{\rm C}}{2} + 1.
    \label{eqn12}
\end{equation}
Using the sampling frequencies at the different stages of processing and the respective filter orders, the starting sample equals 16001 or 1.6001\,s. Hence, the Wiener output is subtracted from the 10\,kHz target data between samples 16001 and 26000. This also implies that the delay introduced in the one second long NN canceled output is 6000 samples or 0.6\,s. Figures \ref{ToyFIR}(a)-(e) show the shifts in the data at each stage of the NNC application.

\section{Proof of principle}
\label{sec:principle}
\begin{figure}
\begin{center}
    \includegraphics[width=0.5\textwidth]{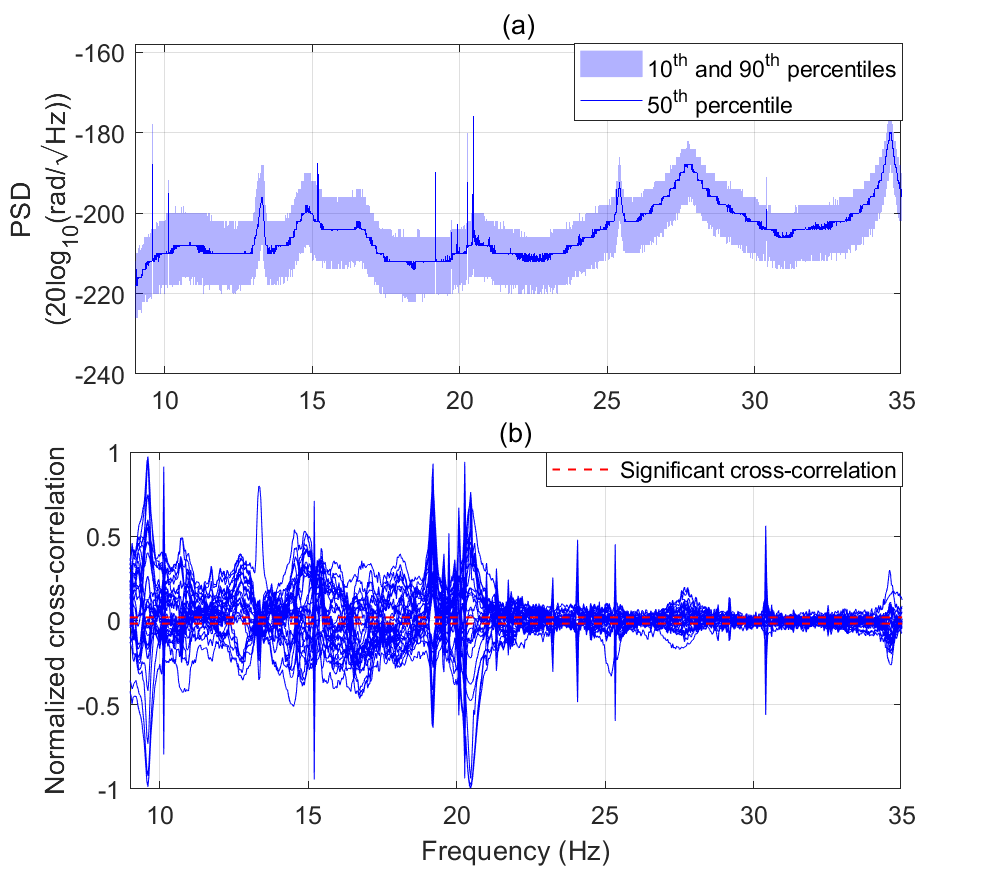}
    \caption{(a) The blue solid line shows the $\mathrm{50^{th}}$ percentile of the PSDs measured by the tiltmeter during the period May 01 -- 07, 2023. The $\mathrm{10^{th}}$, and $\mathrm{90^{th}}$ percentiles are shown with the blue band. (b) Normalized cross-correlations between the 30 geophones at the NEB and the tiltmeter for the frequency band 10 -- 35\,Hz corresponding to May 01, 2023. The red dashed lines show the level of significant cross-correlation.}
    \label{TiltSig}
\end{center}
\end{figure}

The performance of the NNC system was tested by using the tiltmeter signal measured at the NEB as the target data and the geophone signals as the reference data. The location of the tiltmeter inside the NEB is shown with a red star in Figure \ref{NNCArray}(c). The tiltmeter was initially developed within the Archimedes experiment \citep{calloni2014towards} as a beam-balance prototype and essentially functions as a rotational sensor. The resonance frequency of the tiltmeter is about 25\,mHz corresponding to a center of mass positioning within 10 $\mathrm{\mu m}$ of the bending point. It is equipped with two different optical readout systems comprising a Michelson interferometer for higher sensitivities and an auxiliary optical lever capable of handling a larger dynamic range. A detailed description of the tiltmeter and an assessment of its sensitivity in a quiet seismic environment can be found in \citep{calloni2021high} and \citep{allocca2021picoradiant}, respectively. Figure \ref{TiltSig}(a) shows the $\mathrm{10^{th}}$, $\mathrm{50^{th}}$, and $\mathrm{90^{th}}$ percentiles of the PSDs of the tilt signal. These were estimated by dividing the data into 1000\,s long segments and corresponds to the period May 01 -- 08, 2023. The measured tilt in the 10 -- 20\,Hz band  is about $10^{-11}$ -- $10^{-10}\,\mathrm{rad/\sqrt{Hz}}$ and is comparable to that measured at the LIGO Hanford site \citep{CoEA2018a}. The sharp peaks that appear in the PSDs coincide with Rayleigh waves originating from the HVAC system of the NEB, and was previously established in section \ref{sec:site}. Consequently, strong positive or negative cross-correlations between the tiltmeter and geophone signals are observed at these peaks (figure \ref{TiltSig}(b)). The broader peaks centered at 15, 17, 20, 27, and 34\,Hz, show moderate correlations between 0.2 and 0.4.
The frequency-domain cross-correlations shown in figure \ref{TiltSig}(b) are estimated using  equation (\ref{eq:coherence}), and by dividing an entire day of data into 30\,s long segments. Consequently, the minimum value of significant cross-correlation is $1/\sqrt{2880} \approx 1.81\times 10^{-2}$, which is represented with the red dashed line in the figure.

\begin{figure}
\begin{center}
    \includegraphics[width=0.5\textwidth]{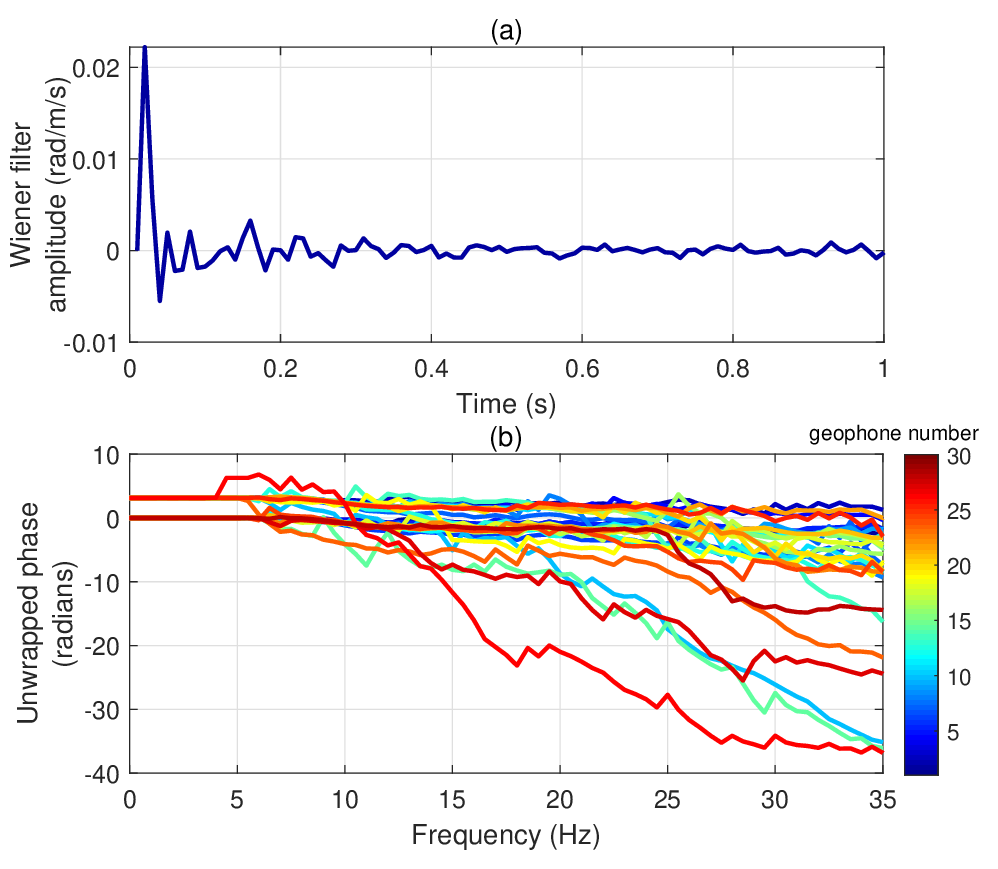}
    \caption{(a) Wiener filter in the time domain for one of the geophone channels. (b) Phase response of the Wiener filters corresponding to all the geophone channels.}
    \label{WFiltTime}
\end{center}
\end{figure}

\begin{figure*}[ht!]
\begin{center}
    \includegraphics[trim={0cm 0.0cm 0cm 0cm},clip,width=\textwidth]{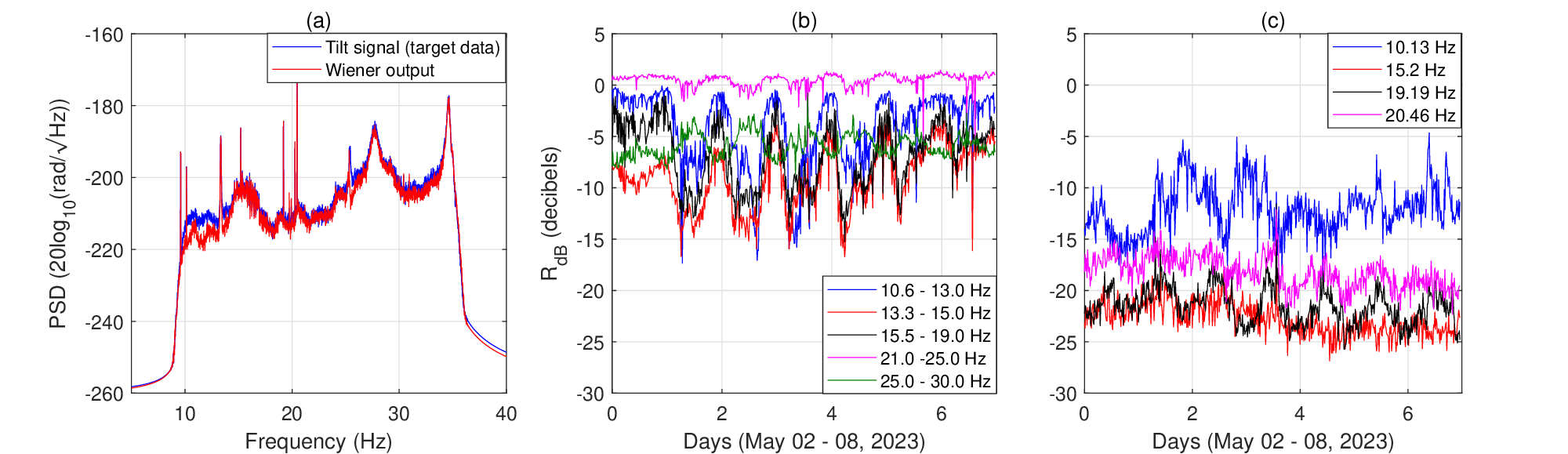}
    \caption{(a) The blue and the red curves show the PSDs of the tiltmeter signal and the signal reconstructed by applying the Wiener filters on the geophone data. (b) Noise cancellation in units of decibels achieved for the tiltmeter signal corresponding to five frequency bands. (c) Noise cancellation in units of decibels achieved for the tiltmeter signal corresponding to four of the sharp spectral peaks.}
    \label{NNCTilt}
\end{center}
\end{figure*}

The first step in the estimation of the Wiener filter for every reference channel is signal preconditioning. Following the steps and the filter orders mentioned in section \ref{sec:wiener}, the tiltmeter signal is first downsampled from 10\,kHz to 100\,Hz, and the geophone signals are downsampled from 500\,Hz to 100\,Hz. Next, the data are high-pass filtered with a passband frequency $\geq$ 10\,Hz. The mean of the 10 -- 35\,Hz signals are then subtracted and the zero-mean signals are used to estimate the cross-correlations between the geophones and that between the tiltmeter and the geophones. We use the data measured on May 01, 2023 to estimate the cross-correlations. Finally, following equation (\ref{eqn9}), the Wiener filter is estimated using all geophones as reference channels. Figure \ref{WFiltTime}(a) shows the amplitudes of the Wiener filter for one of the geophone channels. Filter amplitudes for other geophones look similar. We show the unwrapped phase of filters of all geophones in the frequency domain in figure \ref{WFiltTime}(b). Filters of different geophones have different phase characteristics depending on the geophone locations and capture the propagation characteristics of the seismic noise at different points inside the NEB.
The Wiener filter estimated from one day of cross-correlations (May 01, 2023) is then used to reconstruct the tiltmeter signal for the next seven days (May 02 -- 08, 2023). The blue and red curves in figure \ref{NNCTilt}(a) show the PSDs of the tiltmeter signal and the signal estimated by applying the Wiener filter to the geophone data for a 1000\,s stretch. We denote the error between the measured tilt signal $\tau(k)$ and the reconstructed tilt signal $\hat\tau(k)$ as $e(k) = \tau(k)-\hat\tau(k)$. In the frequency domain, the noise cancellation factor in decibels is then defined as 
\begin{equation}
    \mathcal R_{\rm dB}(f) = 10\times \log_{10}\left(\frac{E(f)}{T(f)}\right)
\end{equation}
where $E(f),\,T(f)$ represent the PSDs of the error signal $e$ and tiltmeter signal $\tau$ at frequency $f$. Since the composition of the seismic field is dependent on the frequency, we further average $\mathcal R_{\rm dB}(f)$ for different frequency bands of interest. Figure \ref{NNCTilt}(b) shows the temporal evolution of the noise cancellation factor for five different frequency bands. These bands are chosen such that they don't overlap with the sharp spectral peaks. 

The best noise cancellation factor of about 10 -- 15\,dB is observed for the bands 13.3 -- 15, 15.5 -- 19, 25 -- 30\,Hz during the day time. The cancellation factor is less than 5\,dB during the night time. It is worth noting that the Wiener filter that is estimated using a day of data is dominated by strong noise cross-correlations that occur during the day, hence a better cancellation is observed during the day time. However, the lack of noise cancellation during the night does not add noise to the subtracted signal. The worst performance is observed for the band 21 -- 25\,Hz, where the noise cancellation factor is slightly above 0\,dB, implying that it adds little noise to the output. Similar to the broadband case, the evolution of the noise cancellation factors for the spectral peaks are shown in figure \ref{NNCTilt}(c). As expected, we observe a better noise cancellation factor of more than 15\,dB and is in accord with the strong Rayleigh wave content of these signals. Unlike the broadband case, little diurnal variation is observed. These signals are characterized by a strong SNR and a stationary phase, and are affected little due to interference from local transient noise sources. 

\begin{figure}[ht!]
\begin{center}
    \includegraphics[trim={0cm 0.0cm 0cm 0cm},clip,width=0.5\textwidth]{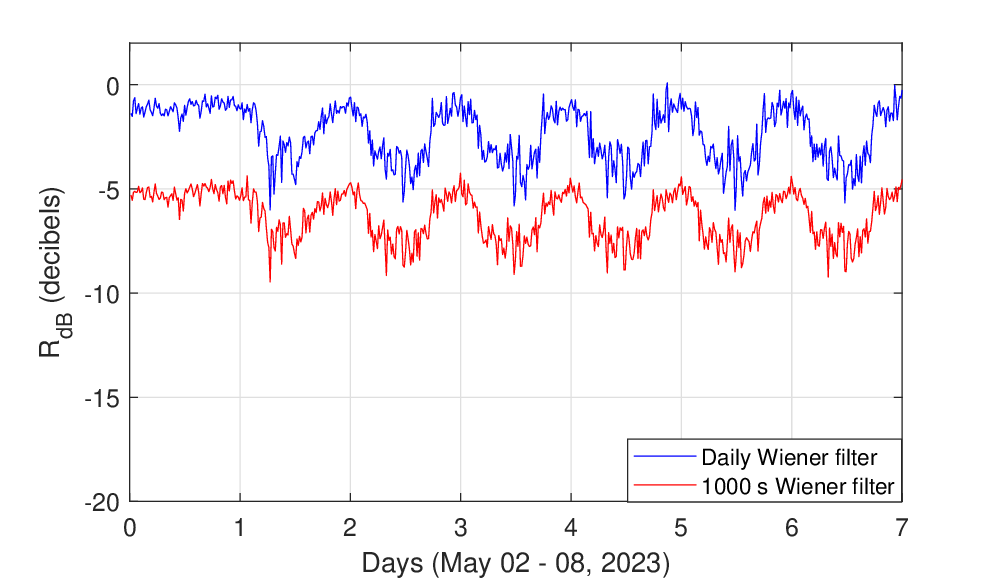}
    \caption{Noise cancellation in units of decibels achieved for the tiltmeter signal for the frequency band 21 -- 25\, Hz corresponding to the two cases: i) when the Wiener filter is updated daily (blue curve), and ii) when the filter is updated every 1000 s (red curve).}
    \label{21_25HzEvol}
\end{center}
\end{figure}

\begin{figure}[ht!]
\begin{center}
    \includegraphics[trim={0cm 0.0cm 0cm 0cm},clip,width=0.5\textwidth]{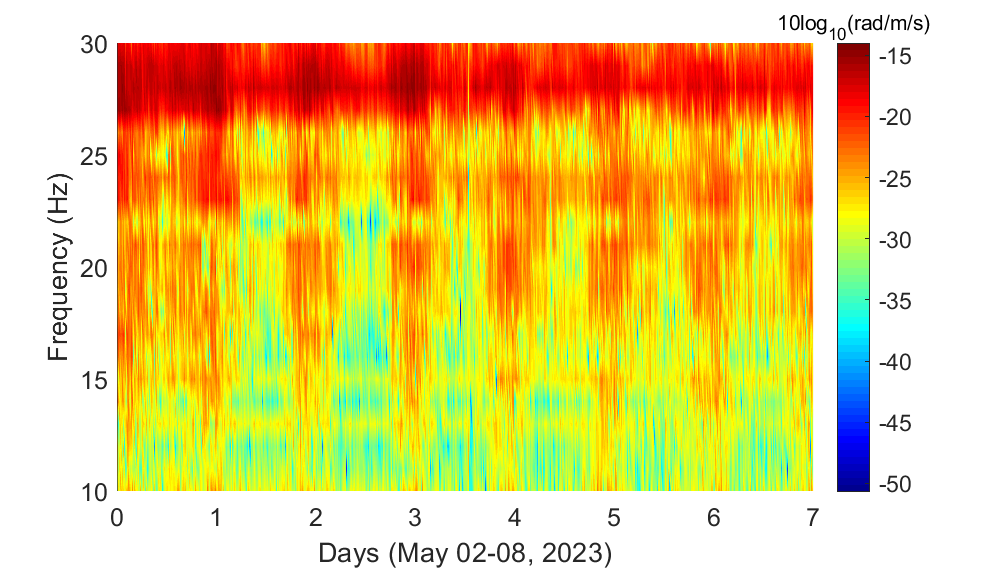}
    \caption{Temporal evolution of magnitude of the Fourier transform of the Wiener filter corresponding to one of the geophones at the NEB.}
    \label{WFiltEvolve}
\end{center}
\end{figure}

The noise cancellation results shown in Figures \ref{NNCTilt}(b) and (c) point to moderate temporal variation in the seismic field characteristics at the site. In particular, the noise cancellation in the band 21 -- 25\,Hz is poor (even enhancing noise). Hence we assessed the performance of the Wiener filter for two cases. In the first case, the Wiener filter was calculated every day and applied to the same day of data. In the second case we calculated the Wiener filter every 1000\,s and applied it to the same data stretch. The blue curve in Figure \ref{21_25HzEvol} shows that no excess noise is added to the output data for the band 21 -- 25\,Hz when the Wiener filter was updated every day. The performance is further improved by about 5\,dB in the case when the filter was updated every 1000\,s. This points to variability in the origin and the propagation characteristics of the noise in this band, and that the static Wiener filter although calculated using a full day of cross-correlations is not optimal. The variability in the direction of propagation of the noise in the 21 -- 25\, Hz band is also observed in Figure \ref{velPhiHist}(b), where the histograms of the estimated direction of propagation does not point to a persistent source of noise. Although the performance of the static Wiener filter for other frequency bands is satisfactory, it must be noted that updating the Wiener filter a few times every day would further improve the cancellation performance. This pattern is also reflected in the temporal evolution of the filter amplitudes for every channel. If no variation in the amplitude and phase characteristic of the filter is observed, that would imply a stationary seismic field and the noise cancellation would not vary with time. We estimate the Wiener filter for the same week during which noise cancellation results were shown earlier. Figure \ref{WFiltEvolve} shows the temporal evolution of the magnitude of the Fourier transform of the Wiener filter corresponding to one of the geophones at the NEB. We observe a diurnal variation with higher amplitudes during the day. Hence, the application of a static Wiener filter which has been estimated using a certain period of the data is not the best solution when the noise varies significantly between days. In such cases the optimal filter needs to be adaptive and should be calculated for every new data sample or data stretch, depending on the needs of the cancellation system. A performance analysis of adaptive Wiener filters is beyond the scope of this work, but such schemes are currently under study and their suitability for a real-time application are being tested. 

\section{Preparations for future performance improvements}
\label{sec:discussion}
A question that needs to be answered as part of the risk management and design phase of the Virgo NNC system is what to do if its performance is not good enough, or not as good as expected. The sensor array, data-acquisition system, and data-processing pipeline are well characterized at this point and function as foreseen. The noise-cancellation performance with the tiltmeter as target channel is as expected, i.e, similar performance was observed at the LIGO Hanford site with a tiltmeter and temporary array deployment \cite{CoEA2018a}. This means that if the performance of the NNC system is not as good as expected, then the most likely explanation is that the sensors do not provide all the required information about environmental fields to do efficient noise cancellation. 

Site characterization, NN modeling, and array optimization were the three most important steps to predict NNC performance. Structural vibrations near the test masses were studied with vibration measurements and NN modeling with the conclusion that they make a small contribution to NN. Arrays of microphones were deployed in all buildings (more than 70 microphones in total). These microphones were planned from the beginning as part of the NNC system to cancel NN from the acoustic field \cite{FiEA2018}. They turned out to be less important to NNC after the detector infrastructure team managed to reduce the level of acoustic noise by making changes to the ventilation system \citep{NEBAHU}.  Only after these changes, Rayleigh-wave NN became the clearly dominant predicted contribution to NN. Nevertheless, the microphones are being used for site characterization, and they might become valuable in the future to improve the NNC performance. Construction of finite-element models has begun for dynamical simulations of the seismic field, implementing all we know about the structure of buildings, surface, and geology. If NNC does not perform as expected, these simulations would provide important information about missed properties of the seismic field and how to adapt the NNC array to improve performance.

Another design modification that promises performance improvements is to switch to time-variant filters, e.g., adaptive Wiener filters. These can take the form of recursive least squares filters, or Kalman filters, etc. We have some indication that correlations of the seismic field change during the day and during the week, and implementing adaptive Wiener filters might improve performance \cite{TrEA2019}. Studies are already underway to explore time-variant filters for noise cancellation and to assess their robustness and effectiveness with respect to static Wiener filters. 

More important design upgrades have been discussed. As can be seen in figure \ref{NNCArray}, the distance of some of the test masses to the building walls is only several meters. The array optimization does not suggest sensor placements outside the buildings, but the calculated optimized arrays are not expected to provide a broadband NN reduction by more than a factor $3$ in amplitude \cite{BaEA2020}. For greater noise reduction, it might be necessary to add outdoor sensors to the array. It will also be investigated whether seismic tiltmeters can improve NNC performance as expected for sites with flat surfaces \cite{HaVe2016}.

Finally, the most advanced design upgrade might come from a robotic sensor array currently under development \cite{BaEA2023}. A pilot project called Flexible Grid Mapping Tool (FGMT) is being carried out at the Virgo interferometer site with the collaboration of the European Gravitational Observatory and the Gran Sasso Science Institute. The FGMT is part of the European research project AHEAD-2020. The idea is to move the array optimization from a simulated environment to the real system. The robots will move the sensor to their optimal locations, and after a data-taking phase at these positions, an improved array configuration will be calculated and the robots will move to their next locations. This process is meant to repeat until the performance of the NNC system converges to its optimum. The main challenges of this system are to manage the robot charging cycles, to navigate with high accuracy inside the buildings, to provide good ground connection of the accelerometers during measurement phases, and to realize a low-latency communication with the Virgo data-acquisition system and timing signal.
 
\section{Conclusion}
\label{sec:conclusion}
In this paper, we have presented the design and implementation of the Newtonian-noise cancellation (NNC) system for the Virgo detector as part of its AdV+ technological upgrades \citep{flaminio2020status}. It is the first such system in the current global network of GW detectors. The main steps that led to the design are (1) selecting sensors, (2) designing the array data-acquisition system, (3) site characterization, (4) Newtonian-noise (NN) modeling, (5) array optimization, (6) design of the noise-cancellation filter, and (7) defining data-processing steps for the online implementation. The design phase started in 2018 and was completed in 2023 soon followed by the completion of the commissioning of the system. 

The AdV+ NNC system consists of 110 vertical geophones whose data are digitized directly at the sensor and transmitted to a central data-acquisition unit at each of the three Virgo stations (two end buildings and the central building). These units communicate with the Virgo data-acquisition system and share its timing signal, which is propagated to all the sensors. More than 70 microphones (the number is increasing steadily due to the interest of the noise-hunting team) have been deployed as well to cancel NN from the acoustic fields. The data of these microphones are not yet included in the online NNC pipeline since acoustic NN is predicted to be a smaller contribution to the total NN.

The first implementation of the NNC pipeline uses a time-invariant, time-domain (FIR) Wiener filter. We studied its performance in a proof-of-principle with a tiltmeter as target channel. We assessed performance limitations and studied their variations with time. The Wiener filter models the PSD of the tiltmeter signal accurately above 15\,Hz, but this does not necessarily mean good coherent noise-cancellation performance. For example, in the 21--25\,Hz band, the cancellation performance diminishes significantly within a few hours after the data stretch used to calculate the filter. In this band, the performance does not get better again later on, which is different from the clear day-night cycle in performance seen in other bands. A careful study of this interesting observation is needed. In any case, it is to be expected that a time-variant Wiener filter will significantly improve the average noise-cancellation performance coming from these temporal changes. 

According to our predictions, the AdV+ NNC design meets the requirements for a factor 3 NN reduction in average \cite{BaEA2020}. The predicted performance depends on a model of seismic NN, which has limitations since the site characterization only produced measurements of vertical surface displacement. These limitations could be overcome by doing new measurements with three-axis seismometers; some of those deployed inside boreholes. Simulations based on refined finite-element models will be important as well for future improvements of NNC performance. The impact of NN transients has not been analyzed yet. While future NN observations might lead to better NNC designs, improving NNC designs beyond state-of-the-art will become an ever more challenging problem. An increasing amount of details concerning geology, topography and more extensive surveys of the seismic field, and possibly other NN contributions from structural vibrations and the atmosphere will have to be considered. The experience of the next years will be crucial to assess the true complexity of NNC also with respect to proposed next-generation detectors like the Einstein Telescope \cite{ET2020} or Cosmic Explorer \cite{EvEA2021}.

\section{Acknowledgements}
The authors gratefully acknowledge the Italian Istituto Nazionale di Fisica Nucleare (INFN), the French Centre National de la Recherche Scientifique (CNRS) and the Netherlands Organization for Scientific Research (NWO), for the construction and operation of the Virgo detector and the creation and support of the EGO consortium. The authors also gratefully acknowledge research support from these agencies as well as by the Spanish Agencia Estatal de Investigaci\'on, the Consellera d’Innovaci\'o, Universitats, Ci\`encia i Societat Digital de la Generalitat Valenciana and the CERCA Programme Generalitat de Catalunya, Spain, the National Science Centre of Poland and the European Union—European Regional Development Fund; Foundation for Polish Science (FNP), the Hungarian Scientific Research Fund (OTKA), the French Lyon Institute of Origins (LIO), the Belgian Fonds de la Recherche Scientifique (FRS-FNRS), Actions de Recherche Concert\'ees (ARC) and Fonds Wetenschappelijk Onderzoek—Vlaanderen (FWO), Belgium, the European Commission. The authors gratefully acknowledge the support of the NSF, STFC, INFN, CNRS and Nikhef for provision of computational resources.\\
Soumen Koley acknowledges the support through a collaboration agreement between Gran Sasso Science Institute and Nikhef and from the European Gravitational Observatory through a collaboration convention on Advanced Virgo +. The authors also gratefully acknowledge the support of the Italian Ministry of Education, University and Research within the PRIN 2017 Research Program Framework, n. 2017SYRTCN. Tomasz Bulik, Marek Cieslar, Mateusz Pietzak, and Mariusz Suchenek are supported by the grant ``AstroCeNT: Particle Astrophysics Science and Technology Centre'' (MAB/2018/7) carried out within the International Research Agendas programme of the Foundation for Polish Science (FNP) financed by the European Union under the European Regional Development Fund. Tomasz Bulik, Marek Cieslar, Mateusz Pietzak, and Mariusz Suchenek are supported by the funding from the European Union’s Horizon 2020 research and innovation programme under grant agreement No 952480 (DarkWave). Tomasz Bulik, Bartosz Idzkowski,  were supported by the TEAM/2016-3/19 grant from the Foundation for Polish Science.
\bibliographystyle{apsrev}
\bibliography{references}

\end{document}